\documentclass[twocolumn,showpacs,preprintnumbers,amsmath,amssymb]{revtex4}
\usepackage{graphicx}% Include figure files
\usepackage{dcolumn}% Align table columns on decimal point
\usepackage{bm}% bold math

\begin{document}
\title{Non-thermal high-energy emissions from black holes by a relativistic capillary effect}
\author{Maurice H.P.M. van Putten}
\affiliation{MIT-LIGO, 175 Albary Street, NW17-161, Cambridge, MA 02139}
\begin{abstract}
Gravitational spin-orbit interactions induce a relativistic capillary effect along 
open magnetic flux-tubes, that join the event horizon of a spinning black hole to infinity.  
It launches a leptonic outflow from electron-positron pairs created near the black hole, which 
terminates in an ultra-relativistic Alfv\'en wave. Upstream to infinity, it maintains a clean 
linear accelerator for baryons picked-up from an ionized ambient environment. We apply it
to the origin of UHECRs and to spectral energy correlations in cosmological gamma-ray 
bursts. The former is identified with the Fermi-level of the black hole event horizon, the 
latter with a correlation $E_pT_{90}^{1/2}\simeq E_\gamma$ in HETE-II and Swift data.
\end{abstract}

\maketitle

\section{Introduction}

Non-thermal high-energy emissions are characteristic properties of active galactic nuclei and
cosmological gamma-ray bursts. These emissions are commonly associated with ultra-relativistic 
baryon-poor outflows, representing the spinning black holes interacting with matter and 
electromagnetic fields. While black holes are described by merely three parameters of mass,
angular momentum and electric charge, their interactions are mediated by the entire
surrounding spacetime subject to radiation boundary conditions on the event horizon
and infinity. For spinning black holes, Mach's principle implies frame-dragging, in which zero-angular 
momentum states assume zero angular velocities relative to the distant stars while corotation with the 
black hole in its close proximity. Even though these two examples -- AGN and GRBs -- represent systems 
of hugely different scales, the inherently scale-free continuum of macroscopic black holes allows scaling 
across a wide range of black hole masses, and hence a common mechanism for producing non-thermal 
high-energy emissions in AGN and GRBs.

The identification of Ultra-High Energy Cosmic Rays (UHECRs) beyond the Greisen-Zatsepin-Kuzmin
cut-off with nearby active galactic nuclei (AGN) poses the question on the physical origin with
their remarkable energies of up to and beyond 1 exa-eV (10$^{18}$eV) \citep{gre66,aug07}.
The observed exa-eV energies probably represent the {\em raw potential energies} generated by the 
inner engine of an AGN, rather than a generally less efficient prompt emission process in, 
e.g., turbulent shocks. As such, the UHECRs point towards a clean site for open, linear 
acceleration, which is largely uncontaminated by surrounding baryonic matter and is not subject 
to in-situ pair-cascade processes.

The association of gamma-ray burst emissions and their afterglows with non-thermal dissipation
of the kinetic energy in ultra-relativistic baryon-poor jets poses the question on the physical 
mechanism producing these outflows \citep{ree92}, presumably out of ab-initio baryon-free 
leptonic outflows following pick-up of baryonic contaminants \citep{she90}. While spinning black 
holes should naturally be active, and pair-creation of leptonic matter is expected to be prominent in
their vicinity, any viable mechanism for launching ultra-relativistic leptonic jets appears to 
be non-trivial in accounting for both the observed macroscopic phenomenology in energies, 
luminosities and durations, and detailed spectral energy correlations \citep{ama02,ghi04,ghi07}.

Spinning black holes provide an attractive alternative for the inner engines of AGN and GRBs
in view of the potentially large amount of spin-energy per unit mass
according to the Kerr metric \citep{ker63}:
\begin{eqnarray}
E_{spin}=\frac{1}{2}\Omega_H^2 I f_s^2,~~f_s=\frac{\cos(\lambda/2)}{\cos(\lambda/4)}.
\label{EQN_ESP}
\end{eqnarray}
In geometrical units in which Newton's constant and the velocity of light are set equal to
1, $\Omega_H=(1/2M)\tan(\lambda/2)$ denotes the angular velocity of the 
black hole of mass $M$, $I=4M^3$ its moment of inertia in the limit of slow rotation 
\citep{tho86}, and $0.7654\le f_s\le 1$ is a moderation associated with relativistic spin 
rates. With $-\pi/2\le \lambda \le\pi/2$, $E_{spin}/M$ can reach a 29\% -- which is one order 
of magnitude larger than the specific spin energy of a neutron star. In view of (\ref{EQN_ESP}), it 
is tempting to associate rotating black holes as the energy source in a wide variety of AGN
and, more generally, in extreme transient sources (recently reviewed in \citep{lev02}).
For example, quasars may represent AGN that harbor a supermassive black hole which spins rapidly. 
The Kerr solution further elucidates some remarkable physical aspects black hole radiation processes.
Perhaps first and foremost, spontaneous emission of a particle by a spinning black hole satisfies 
the Rayleigh criterion 
\begin{eqnarray}
a_p\ge 2M> M\ge a,
\label{EQN_RAL}
\end{eqnarray}
where $a_p$ denotes the specific angular momentum of a radiated particle, $a$ 
denotes the specific angular momentum of the black hole of mass $M$ and radius 
$r_H=2M\cos^2(\lambda/2)$ with spin rapidity $\lambda$, $\sin\lambda=a/M$. 
A specific astrophysical process is required, however, to engage black holes in radiation 
processes by suppressing canonical angular momentum barriers \citep{unr74}, which
otherwise stabilize isolated macroscopic black against spontaneous spin-down.
Most astrophysical processes involve an electromagnetic field, and this offers
some novel prospects for black hole radiation processes according to the second law of 
thermodynamics \citep{car73}
\begin{eqnarray}
\delta M = \Omega_H \delta\left(J_H+J_{em}\right)+T_H\delta S_H + [V]_\infty^H\delta q,
\end{eqnarray}
where $-\delta M$ represents the energy output, $J_{em}$ refers to the angular momentum of the 
electromagnetic field in the presence of a voltage difference $[V]_\infty^H$ between the event 
horizon (with electric charge $q$) and infinity, and $T_H\delta S_H$ denotes the associated
creation of entropy. Recent spectroscopic evidence for rapidly rotating outflows provides
qualitative support for the magnetic fields in disks around supermassive black holes \citep{you07},
consistent with launching outer jets in the form of baryon-rich disk winds by magnetic pressure 
\citep{sto94,van96}.

The Penrose process provides an explicit demonstration that black hole spin can be released to 
infinity \citep{pen69}. However, it is tuned by hand, rather than representing a natural
process, and it is limited to the ergosphere. Nevertheless, it gives a rigorous demonstration 
of principle which satisfies causality and the appropriate ingoing and outgoing radiative
boundary conditions on the event horizon and, respectively, infinity. It hereby
elucidates a ``two-particle" description for the extraction of black hole spin energy.

Recently, we demonstrated the possibility for the emission of positive energy-angular momentum 
particles to infinity along a magnetic flux-tube along the spin-axis of the black hole. 
In an asymptotically uniform magnetic field \citep{note07b}, Hawking radiation is modified according to 
\citep{van00}
\begin{eqnarray}
\frac{d^2N}{dtdN}=\frac{\Gamma}{e^{2\pi(E-V_F)/g_H}+1},
\label{EQN_H}
\end{eqnarray}
where $\Gamma$ denotes the gray-body factor, $E$ the energy of the radiation particle
and $V_F$ denotes the horizon Fermi-level
\begin{eqnarray}
\nu\Omega_H=e\mbox{EMF}_\nu
\label{EQN_NU}
\end{eqnarray}
associated with the angular momentum $\nu=\pm eA_\phi$ of particles with charge $\mp e$ along 
a magnetic flux-surface $2\pi A_\phi$.
It demonstrates that black hole radiation processes are inherently non-local (on the macroscopic
scale set by the size of the black hole), here comprising frame-dragging subject to the radiation 
conditions on the event horizon and infinity. The equivalent integral to (\ref{EQN_H}) is the net 
electromagnetic force EMF$_\nu$ in (\ref{EQN_NU}), associated with a loop which extends and closes 
over infinity, the spin-axis, the horizon surface and the flux-surface $\nu$ at hand.

The result (\ref{EQN_NU}) shows that angular momentum barriers can be successfully circumvented by 
gravitational spin-orbit coupling through an equivalent level-shift \citep{van05}
\begin{eqnarray}
{\cal E}=\omega J
\label{EQN_E1}
\end{eqnarray}
of particles with angular momentum $J=-\gamma g_{\phi\phi}(\Omega-\omega)$, where $\Omega$ denotes the
angular velocity relative to infinity, $\omega$ is the local angular velocity  
frame-dragging in the metric $g_{ab}$, and $\gamma$ denotes the time component $u^t$ of a velocity 
four-vector $u^b$ $(u^cu_c=-1)$. 
Frame-dragging assumes a maximal magnitude on the event horizon, where $\omega=\Omega_H$. 
It decays algebraically according to $r^{-3}$ upon approaching infinity, where $r$ denotes the radial coordinate 
in Boyer-Lindquist coordinates. Thus, frame-dragging is differential in nature, i.e., $\partial_r\omega$ 
is strong near the black hole and weak at larger distances. It is hereby not a gauge-effect.
Two particles with orbits at locations $r_{1,2}$ on a common flux-surface which share the same angular 
velocity $\Omega$, experience a potential difference
\begin{eqnarray}
[{\cal E}] = -\gamma g_{\phi\phi}\left[\omega(\Omega-\omega)\right]~~~~~~~~~~~~~~~~~~~~~~~~~~~~~~\\
 =\left\{\begin{array}{ll}
 -\gamma g_{\phi\phi} [\omega] (\Omega -(\omega)) & (r_2-r_1<<r_2+r_1)\\
 \gamma g_{\phi\phi} \omega_1\left(\Omega-\omega_1\right) & (r_2>>r_H)
\label{EQN_E2}
\end{array}\right.,
\end{eqnarray}
where $[f]=f(r_2)-f(r_1)$ and $(f)=f(r_2)+f(r_1)$.
The results (\ref{EQN_E1}-\ref{EQN_E2}) are universal: they hold regardless of the origin 
of $J$ or $\Omega$, e.g., apply to mechanical or electromagnetical angular momentum.

In (\ref{EQN_H}), electron-positrons have $J=\pm eA_\phi$ \citep{note07}. The result (\ref{EQN_E1})
reveals positive and negative energy orbits exist along the spin axis of a black hole of particles 
in (radiative) Landau states along a magnetic flux-tube. It is dramatically distinct from the ergo-sphere, 
where all circular orbits (of uncharged particles) have positive energy \citep{note07c}. 
An open magnetic flux-tube along the spin-axis of the black hole, may now feature pair-wise 
radiation of positive and negative energy and angular momentum to infinity and, respectively, into the 
black hole--analogous to the aforementioned Penrose process in the ergo-sphere. For the ingoing radiation,
see further \citep{lev02}.

In this {\em Letter}, we point out that the ab-inito interaction (\ref{EQN_H})-(\ref{EQN_E1}) creates 
an open linear accelerator {\em upstream of an ultra-relativistic Alfv\'en front}, whose characteristic
energy is consistent with UHECRs. The resulting emission of UHECRs is accompanied by simultaneous absorption 
of negative energy and angular momentum particles by the black hole. We apply our model to the recent 
identification of UHECRs with AGN and to some recent spectral energy correlations in GRBs.

The gradient (\ref{EQN_E1}) with respect to $r$ is commonly interpreted in terms of the Lorentz invariant 
$E\cdot B$, which is non-zero in a vacuum Wald field \citep{wal74}. It reflects the spin-induced
non-zero curl of the electric field in Boyer-Lindquist coordinates (observers that are not rotating 
relative to the distant stars) \citep{tho86,van00}. Its integral is EMF$_\nu$ in (\ref{EQN_NU}). It 
naturally induces a tendency for charge-separation, until the invariant $E\cdot B$ is small, especially 
so in regions with an effective pair-cascade process \citep{bla77}. The EMF$_\nu$ is invariant 
under any such charge-separation in the zero-dissipation and force-free limit, as charge-separation merely 
produces a local redistribution of electric fields in a semi-infinite open magnetic flux-tube. 

The above shows that a spin-induced $E\cdot B\ne0$ gives rise to a {\em relativistic capillary effect}: 
an open magnetic flux-tube will rapidly develop a continuously extending, largely force-free 
($E\cdot B\simeq 0$) section with uniform angular velocity (upon furthermore neglecting the 
inertia of the charged particles away from the event horizon \citep{gol69,bla77,tho86}). It is bounded 
by two freely moving boundaries in the form of two ultra-relativistic Alfv\'en fronts (a double-transonic 
flow \cite{tak90}). One moves outwards to infinity, and one moves inwards to the event horizon of the black 
hole. This pair of Aflv\'en surfaces represents two corotating Faraday disks \citep{tak90,van01,pun03,lev02}. While the
ingoing Alfv\'en surface never reaches the event horizon in a force-free state \citep{ruf75,pun03,lev02}, 
the corotating outgoing Alfv\'en surface does communicate a near-horizon induction voltage out to a larger 
distance. The outgoing Alfv\'en surface will continue to moves outwards to infinity {in a continuous effort} 
to reduce to $E\cdot B\simeq0$ along the semi-infinite flux-tube--a black hole spin-induced 
capillary effect. While near the black hole, copious production of $e^\pm$ is expected,
canonical cascade processes are insufficient to produce in-situ pair-creation upstream of the outer 
Alfv\'en surface. Here, the magnetic flux-tube remains in a Wald vacuum state, until the 
outgoing Alfv\'en wave passes by--a relativistic capillary motion.

Charge-separation, by capillary extraction of pairs from the pair-rich black hole environment,
introduces a finite section between $r_2$ and $r_1$ which, if force-free and highly conductive, will
have zero electrostatic potential difference as seen in Boyer-Lindquist coordinates. Because the
EMF$_\nu$ defined above is unchanged under charge-separation (it performs no work), the 
potential difference (\ref{EQN_E2}) now emerges {\em outside} of the force-free section between 
$r_2$ and $r_1$. In what follows, we identify $r_1$ and $r_2$ with the aforementioned (time-dependent) 
Alfv\'en surfaces and denote their common angular velocity by $\Omega_A$.

The Alfv\'en surfaces communicate a potential difference (\ref{EQN_E2}) along the semi-infinite
region $[r_2,\infty]$ upstream of the outgoing surface and the finite region $[r_H,r_1]$ 
upstream of the ingoing surface. This state is astrophysically realistic, provided there is an 
external source of the magnetic field, such as may be supported by an ionized medium surrounding the
black hole in the form of an accretion disk or torus. It then predicts (a) the formation of
an outgoing leptonic outflow represented by region enclosed by the two Alfv\'en surfaces,
(b) a linear accelerator upstream of the outgoing Alfv\'en front and (c) a characteristic
energy scale given by the Fermi-level of the event horizon of the black hole.

We apply the relativistic capillary effect to the origin of UHECRs from AGN and ultra-relativistic 
outflows from the inner engine to cosmological gamma-ray bursts.

The Fermi-level in (\ref{EQN_H}) supports, by aforementioned corotating Alfv\'en surfaces, a  
potential input to the largely force-free region {\em upstream} of the outgoing Alfv\'en surface. It 
hereby defines, to within a factor of order unity associated with current-induced potential drops, 
the generator voltage across the linear accelerator which extends to infinity upon taking the limit 
$r_1\rightarrow r_H$ in (\ref{EQN_E2}),
\begin{eqnarray}
[{\cal E}] =-\gamma g_{\phi\phi} \left(\Omega_A-\Omega_H\right)\Omega_H.
\end{eqnarray}
When the outer Alfv\'en surface has moved out to a large distance, $[{\cal E}]$ represents the generator
potential in the linear accelerator upstream, while the sum of $[{\cal E}]$ plus the potential difference 
between the inner Alfv\'en surface and the event horizon equals $V_F$, apart from aforementioned
current-induced potential drops across the event horizon. The latter represents dissipation,
according to \citep{van01}
%\begin{eqnarray}
$T_H\dot{S}_H=\frac{1}{2}||\dot{A}_\phi||^2+2||I_H||^2.$
%\label{EQN_D}
%\end{eqnarray}
In the limit of slow variations, it reduces to the horizon dissipation given by the impedance of 
vacuum \citep{tho86} and associated with Maxwell stresses on the event horizon \citep{ruf75}.

Thus, baryons picked up from an ionized environment in upstream of the outgoing
Alfv\'en surface receive energies of the order of $V_F$ whenever $\Omega_A$ is between
0 and $\Omega_H$ (the same condition as in \citep{bla77,tho86}), i.e.,
\begin{eqnarray}
{\cal E}=\left(\frac{M}{10^9M_\odot}\right)
  \left(\frac{B}{10^4\mbox{G}}\right)\left(\frac{\theta_H}{0.1}\right)^2\mbox{exa-eV}
\label{EQN_VF}
\end{eqnarray}
where exa=$10^{18}$ and $\theta_H$ denotes the half-opening angle of the open magnetic flux-tube on the event horizon of 
the black hole. 

The high-energy emissions from gamma-ray bursts have been attributed to the dissipation of kinetic energy
in ultra-relativistic baryon-poor jets. Let $c_1$ denote the ratio of observed peak energy $E_p$ 
to ${\cal E}$, comprising the combined efficiency of converting ${\cal E}$ into kinetic energy in a neutron-enriched 
leptonic jet with subsequent conversion into high-energy gamma-rays in internal and external shocks. 
The associated luminosity in Poynting flux along the open magnetic flux-tube in the force-free 
limit \citep{gol69,bla77,tho86}
\begin{eqnarray}
L=\Omega^2_AA_\phi^2,
\label{EQN_L}
\end{eqnarray}
Let $c_2$ denote the efficiency 
of converting $L$ into the true energy in gamma-rays, i.e., $E_{\gamma}=c_2LT_{90}$. Then (\ref{EQN_E1}) 
predicts a positive correlation between peak-energies in gamma-rays and the true energy in gamma-rays: 
$E_pT_{90}^{1/2} = ek E_{\gamma}^{1/2}$, where $k=2c_1/\sqrt{c_2}$. Assuming $c_2\propto c_1$ and in the 
approximation of $\Omega_F,\omega$ to be of order $\Omega_H$, the Ghirlanda relation \citep{ghi04}
\begin{eqnarray}
E_p\propto E_\gamma^{0.7}
\end{eqnarray}
together with the Eichler \& Jontof-Hutter correlation between peak-energies and kinetic energy of the 
outflow \citep{eic05}
%\begin{eqnarray}
$c_1\propto E_p^{3/2}$
%\end{eqnarray}
gives $k\propto E_{\gamma}^{21/40}\simeq E_{\gamma}^{1/2}$. It follows that
\begin{eqnarray}
E_pT_{90}^{1/2} \propto E_{\gamma}.
\label{EQN_C1}
\end{eqnarray}
The correlation (\ref{EQN_C1}) bears out remarkably well the current HETE-II and Swift data in 
\citep{ama02,ghi04,ghi07}, compiled 
in Fig. 1. It should be mentioned that alternative explanations on the basis of viewing angles 
are also consistent the Ghirlanda relationship \citep{ghi04}.
\begin{figure}
\centerline{\includegraphics[angle=00,scale=.4]{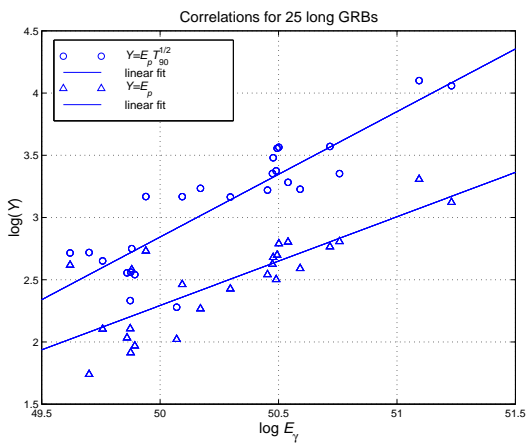}}
\caption{Shown is the correlation in the published data of \citep{ama02,ghi04,ghi07} between the
 peak energy $E_p$, the true energy $E_\gamma$ 
 in gamma-rays of long-duration gamma-ray bursts 
 with known redshifts and inferred opening angles, assuming a stellar-wind type environment 
 to the gamma-ray burst progenitor and using redshift corrected durations $T_{90}$. 
 A linear fit gives a slope and Pearson coefficient 
 $(s,c)=(1.01,0.85)$ to $Y=\log\left(E_pT_{90}^{1/2}\right)$ 
 ({\em circles}) and $(s,c)=(0.71,0.76)$ in the Ghirlanda correlation to 
 $Y=\log\left(E_p\right)$ ({\em triangles}). \label{fs1}}
\end{figure}

To summarize, a frame-dragging induced capillary effect in an open magnetic flux-tube links the
event horizon of a spinning black hole to infinity which predicts
a leptonic outflow downstream of an outgoing Alfv\'en surface and a linear accelerator upstream.
The former is expected to be largely force-free and Poynting-flux dominated as envisioned 
in \citep{bla77}. It serves to communicate the horizon Fermi-level to the the boundary of the latter,
which itself remains largely charge-free as pair-production by spontaneous emission and/or canonical 
cascade processes are limited to the region around the black hole. The linear accelerator 
hereby will accelerate protons from an ionized environment up to the raw Faraday-induced potentials
of spinning supermassive black holes, consistent with the observed UHECRs in AGN. 
It predicts repeat events in \citep{aug07} within about 100 years at current detection statistics. 
The leptonic outflow itself predicts a correlation between peak-energies, durations and true energies 
in gamma-rays (corrected for geometric beaming) in cosmological gamma-ray bursts from rapidly spinning 
black holes consistent with recent HETE-II and Swift data.

{\bf Acknowledgment} The author thanks the hospitality of the Physics Department of Nanjing University,
where some of the work was performed.


\begin{thebibliography}{}
\bibitem[note07b(2007)]{note07b} 
 We consider the black hole to be in its lowest energy state, whereby it assumes maximal horizon flux 
 of magnetic field at all spin-rates. The lowest energy state is described by a Wald charge $q_H=2BJ_H$, 
 where $B$ denotes the strength of the magnetic field and $J_H$ the angular momentum of the black hole
 \citep{wal74}.
\bibitem[note07c(2007)]{note07c}
  Explicit solutions to the specific energy $e=e(r)$ and specific angular momentum $j=j(r)$ of 
  circular orbits in the equatorial plane have been derived by \citep{bar72}. Their minima at
  $e^\prime(r)=0$ and $j^\prime(r)=0$ coincide, and define the inner most stable circular orbit.
  The explicit solutions show that the minimum $e(r)$ at $e^\prime(r)=0$ is always positive. 
  The Penrose process in the ergo-sphere hereby necessarily involves non-circular trajectories.
\bibitem[note1(2007)]{note07}
 Since $e AB/2\pi=eA_\phi = J = m_e (\omega/2\pi) A$, where $A$ denotes the cross-sectional area of the 
 flux-surface with enclosed flux $2\pi A B$, the magnetic moment $\mu=-(e/m_e)J$ gives
 $E=\omega J\equiv-\mu B$. It forms the integral of both the
 invariant $E\cdot B$ in the Wald solution. As shown in \citep{van05}, it also represents the 
 integral of the local interaction of the Riemann tensor with angular momentum, as discussed in 
 \citep{pap51} and \citep{tho86}.
\end{thebibliography}
\end{document}